\documentclass[english]{article} 
\usepackage[T1]{fontenc} 
\usepackage[latin1]{inputenc} 
\usepackage{graphicx}
\usepackage{babel} 
\makeatother 
\begin{document}
\title{The accreting neutron stars are quasars, and the universe does not
expand.
\author{Jacques Moret-Bailly 
\footnote{Laboratoire de physique, Universit\'{e} de Bourgogne, BP 47870,
F-21078
Dijon cedex, France. email : Jacques.Moret-Bailly@u-bourgogne.fr%
}}}

\maketitle
\begin{abstract} The reliable theory of the evolution of heavy stars predicts
the existence of a type of neutron stars which accrete a cloud of dirty
hydrogen (accretors). Although they are very small (some hundreds of
kilometres), the accretors should be easily observable because the accretion
raises the surface temperature over 1 000 000 K, but they are never detected.

The reason of this failure is a misunderstanding of the spectroscopy of
hydrogen crossed by a powerful beam of short wavelengths light.

Except very close to the surface, hydrogen is mostly heated by a Lyman
absorption improved by a parametric frequency shift due to excited atomic
hydrogen, so that this absorption stabilises the temperature between the
limits of ionisation and dimerisation. A powerful radio emission may produce
an extra ionisation where the pressure allows a good electrical conduction.

The combination of Lyman absorptions and redshifts produces an instability
which chains Lyman absorption patterns: when a redshifted Lyman absorbed line
coincides with an other Lyman line of the gas, all absorption lines of the gas
are written into the spectrum.

Thus all characteristics of the complex spectrum of a quasar are generated, so
that observed accretors are named quasars, and the origin of the intrinsic
redshifts is found.

The lack of redshifts of the variations of luminosity of stars and quasars
shows that the \textquotedblleft cosmological redshifts\textquotedblright {}
result from the parametric frequency shift, so that the universe does not
expand.
\end{abstract}

\section*{Introduction}
The parametric light-matter interactions play a big role in laser and microwave
technologies, allowing, for instance, to add, multiply or split frequencies.
These interactions are space-coherent, so that they do not blur the images,
and they do not change the states of the involved molecules. Although the
refraction is a parametric effect, and all interactions start by a parametric
effect, the fugacity of the parametric exchanges of energy with matter leads
to neglect these effects using usual incoherent light. Therefore, it appears
useful to explain simply in the next section, without the trivial
computations, the parametric effects which produce frequency shifts, in
particular the {}``Coherent Raman Effect on Incoherent Light'' (CREIL)
\cite{Mor98a,Mor98b,Mor01,Mor03,Jensen}. 

Hydrogen is the main component of the Universe, and the spectroscopy of small
amounts of this gas is well known. But, in the Universe, long paths allow the
observation of effects which are forbidden at usual pressures. In particular,
strong CREIL frequency shifts are produced by atomic hydrogen if its principal
quantum number\textrm{n}  is low, but larger than 1. Section \ref{chaine}
shows that this property induces an instability which produces a forest of
absorption lines.

Section \ref{quasar} studies the variations of pressure and temperature in a
cloud surrounding a small, heavy, extremely hot object, and the consequences
of these variations on the spectrum emitted by the system.

Section \ref{autres} shows that a lot of observations is more easily understood
using CREIL than using the standard theory.

\section{From refraction to other parametric light-matter
interactions.}\label{refraction}
\subsection{Recall of the analytic theory of refraction.}\label{refrac}
Huygens explained the propagation of the light in the vacuum (fig. 1)
supposing that \emph{all} points of a wave surface A scatter the light
coherently that is are sources of wavelets whose envelope is a new wave 
surface.
\begin{figure} \begin{center} \includegraphics[height=6.5 cm]
{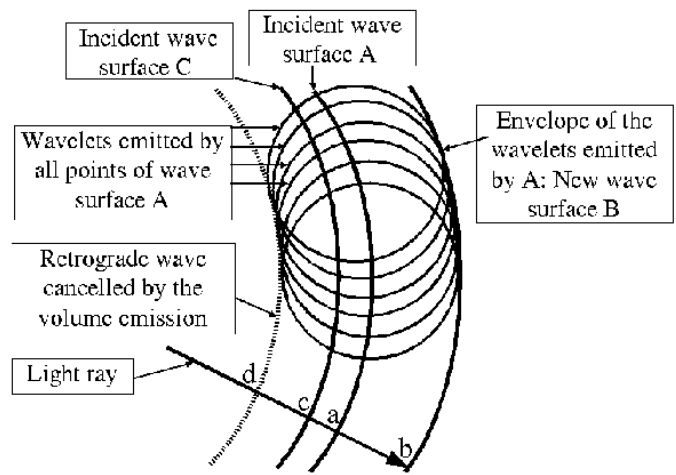}\end{center}
 \caption{ Huygens' construction. } \end{figure}
The coherence of the scattering requires that all points on a wave surface
radiate with the same phase, and here this phase is supposed equal to the 
phase of the incident wave. A retrograde wave is eliminated taking into account
the volume scattering: the paths from the source to point a or an 
other scattering point c, plus the path to b are equal, while they differ to d,
producing, in the volume of the scattering, a cancellation for a backward 
propagation. 

In matter, the molecules\footnote{ This word is used for mono- or polyatomic
molecules, and, more generally for any set of atoms able to scatter 
the light.} scatter the light, (fig. 2 ),
\begin{figure} \begin{center} 
\includegraphics[height=7 cm]{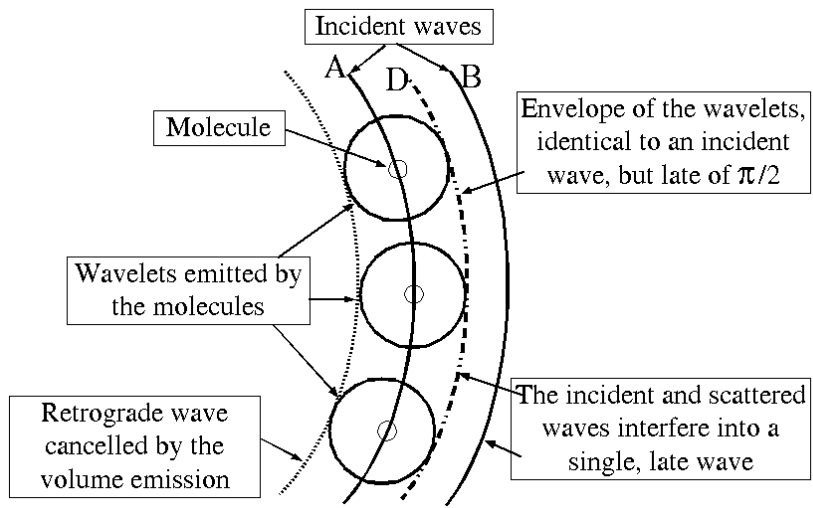}\end{center}
 \caption{ Scattering of light by molecules.}  \end{figure}
 so that, provided that all molecules produce the same phase shift, a Huygens'
construction may be added to the regular one, producing a second 
wave surface D. However the number of molecules is finite, Huygens construction
is not perfect, so that it exists an incoherent scattering, making, 
for instance, the blue of the sky. In the whole paper this incoherent
scattering is neglected.

If the medium is transparent, the scattered wave is late of pi/2. As the wave
surfaces are identical for the incident and scattered waves which have 
the same frequency, the waves interfere into a single, late, refracted wave.

The emission of the scattered wavelets and wave surface D requires a dynamical
excitation of the molecules whose amplitude must be proportional 
to the exciting amplitude to obtain an index of refraction independent on the
intensity. Thus, an energy proportional to this intensity is absorbed by the
refracting molecules, and, as we have assumed that the medium is transparent,
this energy is returned coherently to the wave (fig. 3).
\begin{figure} \begin{center} \includegraphics[height=6 cm]
{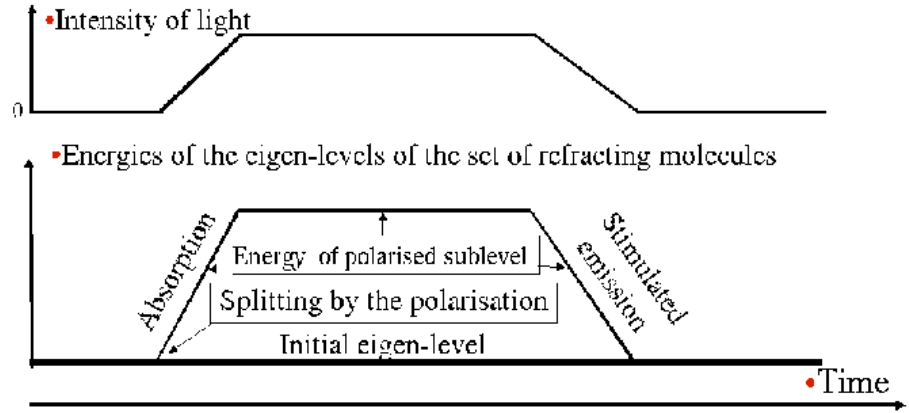}\end{center}
\caption{ Refraction of a pulse of light.} \end{figure}

\subsection{Global, quantum theory of parametric interactions.}
Remark that only a part of an incident energy $h\nu$ is shared among all
molecules of a prism, but that this sub-quantum splitting of the energy is 
allowed by quantum mechanics, the mode of the light beam and the prism making a
single system. In this this representation, there is not a virtual scattering
of the light followed by an interference with the incident beam, but a
transformation of the beam.

Suppose that all $N$molecules of the refracting medium are identical and in the
same, nondegenerate state (else, the effects add). In the dark, the 
degeneracy of the set of $N$ molecules is $N$. A light beam perturbs the
degenerate state, mixing it with other states, breaking the degeneracy. The
polarisation state which appears, having got energy from the light beam, and
able to return it, is characterised by a quantum index which is the mode of
the light beam. Remark, on figure 3, that the energy of the polarisation state
depends on the intensity of the light beam.

If several modes interact with the molecules, several states of polarisation
appear. A parametric interaction may perturb the states, but must not destroy
them, preserving the geometry of the modes in an homogeneous medium and the
stationary states of the molecules after the interaction.

\section{The \textquotedblleft Coherent Raman Effect on Incoherent
Light\textquotedblright {} (CREIL).}

\subsection{A simple parametric interaction.}
In the Coherent Raman Effect on Incoherent Light\textquotedblright {} (CREIL),
the interaction is a simple transfer of energy between sublevels of
polarisation (figure 4). As these sublevels have the same parity, this
interaction requires an intermediate state.

\begin{figure} \begin{center} \includegraphics[height=6 cm]
{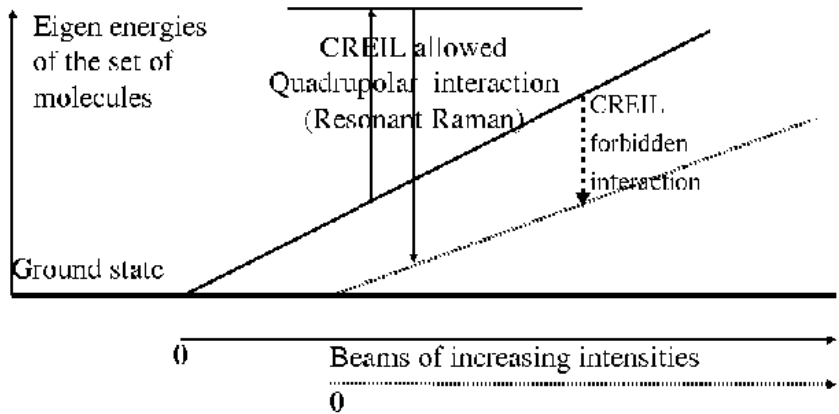}\end{center}
\caption{ The CREIL interaction.} \end{figure}

The interaction must obey the second law of thermodynamics; in the CREIL, it
increases the entropy by a flux of energy from the light modes which have the
highest Planck's temperature to the modes which have the lowest one. An
increase (resp. decrease) of temperature produces an increase (resp. decrease)
of frequency. 

The global theory explains simply the optical mechanism of the parametric
interaction, but a precise study is easier using the analytical representation
in which the resonances correspond to virtual transitions (fig 5) and produce
a scattering (fig 2).

\begin{figure} \begin{center} \includegraphics[height=7.5 cm] {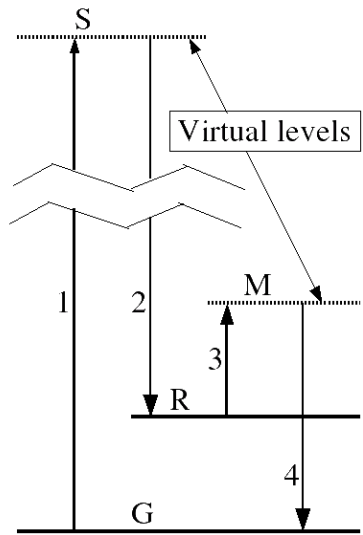}
\end{center}
 \caption{ A standard representation of the CREIL interaction. } \end{figure}

The virtual transitions 1-2 correspond to a virtual Raman effect, transitions
3-4 to a second virtual Raman effect, both Raman effect being simultaneous.

More generally, several beams interact.
\subsection{Conditions for a CREIL effect.}\label{creil}
The refraction and the CREIL are the simplest examples of the parametric
effects. To avoid a destruction of the states, in particular of the modes of
the light beams, two conditions must be verified:

Condition A : The interactions do not change the stationary state of the
molecules, the changes of energy of the molecules being transitory, bound to
the propagation of the light. 

Condition B : The interactions are space coherent, so that they do not blur the
wave surfaces and the images deduced from the incident beams.

G. L. Lamb gives a general condition for a parametric effect \cite{Lamb} :
``The length of the light pulses must be shorter than all relevant time
constants''. 

The parametric effects are widely used with microwave and laser sources, often
with tricks which allow to overcome Lamb's conditions, allowing to add,
subtract, multiply, split the frequencies of beams; using ordinary incoherent
light appears difficult, with the exception of refraction evidently.

\medskip
For CREIL, there are two {}``relevant time constants'', which split condition
B:

Condition B1 : The collisions must not introduce phase shifts as they restart
the scattering, except maybe during short times of collision; a restart is
unimportant in the refraction, but not in effects for which a difference
between scattered and exciting frequencies introduces a phaseshift which
increases with the time. The collisional time which depends on the pressure,
the temperature and the nature of the gas is a {}``relevant time constant''.

Condition B2 : The sum of two sine functions having different frequencies and
the same initial phase is nearly a single frequency sine function whose
intermediate frequency is in proportion of the amplitudes, if the time of
observation is too short to allow the appearance of beats. This mathematical
property is verified using laser beams or a Michelson interferometer having a
slowly moving mirror. Usual incoherent light may be considered as made of
pulses whose length is of the order of a few nanoseconds, so that, to get a
single frequency-shifted wave, the period of the quadrupolar (Raman) resonance
between the levels G and R (fig. 5) must not be shorter than the {}``relevant
time constant'' of the order of 2 nanoseconds, corresponding to a frequency of
500 MHz.

The addition of the sine functions into a single one is approximate, leaving a
residual parasitic wave of different frequency which propagates with a
difference of index of refraction $\Delta n$ due to the dispersion of the
refraction. The corresponding waves radiated at points distant of $L$ along a
light ray have a phase shift $2\pi L \Delta n {} / \lambda$, where $\lambda$
is the vacuum wavelength; when this shift reaches $\pi$, these waves cancel,
so that the scattered parasitic amplitude remains negligible while the
frequency shifts add all along the path.

 \subsection{Intensity of the CREIL effect.}

A precise computation of the intensity of a CREIL effect may be done using
tensors of polarisability which are often not known. A rough, but general
order of magnitude may be deduced from figures 4 and 5 :

Generally the CREIL transfers energy from the hot modes which are high
frequency (infrared, visible, ultraviolet) to cold modes which are in the
thermal radiation. As the quadrupolar resonance ( between levels R and G )
corresponds to a low energy, the three levels R, G, M are close, so that the
corresponding virtual Raman effect is resonant, intense, it does not limit the
intensity of the CREIL effect. Thus the amplitude of the scattering which
produces the CREIL is close to the amplitude of the other coherent Raman
effect, much larger than the amplitude of an incoherent Raman effect.
Therefore, in despite of the low frequency of the quadrupolar resonance, the
CREIL is not a very small effect.

If the S level is low, the CREIL effect is fully resonant, strong. Therefore, a
CREIL effect inside the low energy radiations leads to a fast thermal
equilibrium, a blackbody spectrum for these radiations.
 \section{Absorption of a continuous, high frequency spectrum by low-pressure
atomic hydrogen.}\label{chaine}
In its ground state (principal quantum number $n=1$) atomic hydrogen has the
well known spin quadrupolar resonance at 1420 MHz, too large to provide
frequency shifts by CREIL. In the $n=2$ states, the resonances corresponding
to the quadrupole allowed transitions ($\Delta F = 1$) have the following
frequencies: 178 MHz in the 2S$_{1/2}$ state, 59 MHz in 2P$_{1/2}$ state, and
24 MHz in 2P$_{3/2}$. These frequencies are low enough to allow CREIL, and
high enough to produce a strong CREIL effect. The higher states are generally
less populated, their quadrupolar resonances have lower frequencies, in a
first approximation, we may suppose that only the states $n=2$ or 3 are active
in CREIL.

\medskip
The decay of the states excited by Lyman absorptions heats the gas; if the
intensity of the light is large, the absorption is strong, many atoms are
ionised, do not absorb anymore, so that the heating is limited, the
temperature of the gas stabilises at a value which depends on the intensity in
the Lyman region. Remark that a frequency shift renews the intensity of the
light at the Lyman frequencies, so that the stabilisation of the temperature
works over large distances.

If the temperature is high enough to populate the excited states, the shifting
is permanent, the lines get the width of the redshift, so large that they
cannot be observed.

Suppose now that the temperature is relatively low (10 000 K), so that the
excited levels are populated by Lyman absorptions only(fig 6).
\begin{figure} \begin{center} \includegraphics[height=5 cm] {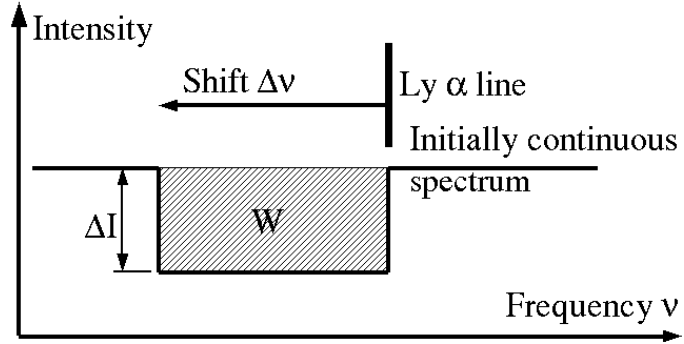}
\end{center}
 \caption{ Absorption by a single line.} \end{figure}

Considering the absorption by the Ly$_\alpha$ line, in an homogenous gas, the
population in the excited state $n=2$ is constant, so that the redshift is
constant, the absorption too (fig. 6). The absorbed energy is proportional to
the product W  of the absorbed intensity $\Delta I$ by the width $\Delta\nu$
of the absorbed line (neglecting the natural width of the line compared with
$\Delta\nu$). Supposing a constant decay of the excited level, the number of
excited atoms is proportional to $W=\Delta\nu\Delta I$. But the redshift
$\Delta\nu$ is proportional to the number of excited atoms, so that $\Delta I$
does not depend on the incident intensity of light.

\begin{figure} \begin{center} \includegraphics[height=5 cm] {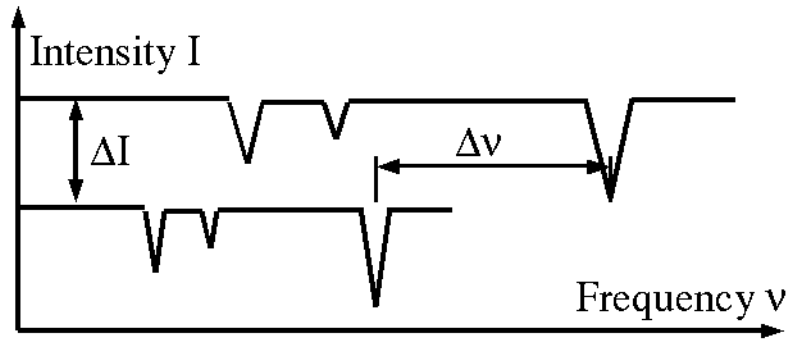}
\end{center}
 \caption{ Absorption of a spectrum by a Lyman line and redshift.}
 \end{figure}

Fig 7 shows the result of the absorption of a spectrum by the Lyman $\alpha$
line: the contrast is increased by the constant absorption while the scale of
frequencies is changed.
\begin{figure} \begin{center} \includegraphics[height=10 cm] {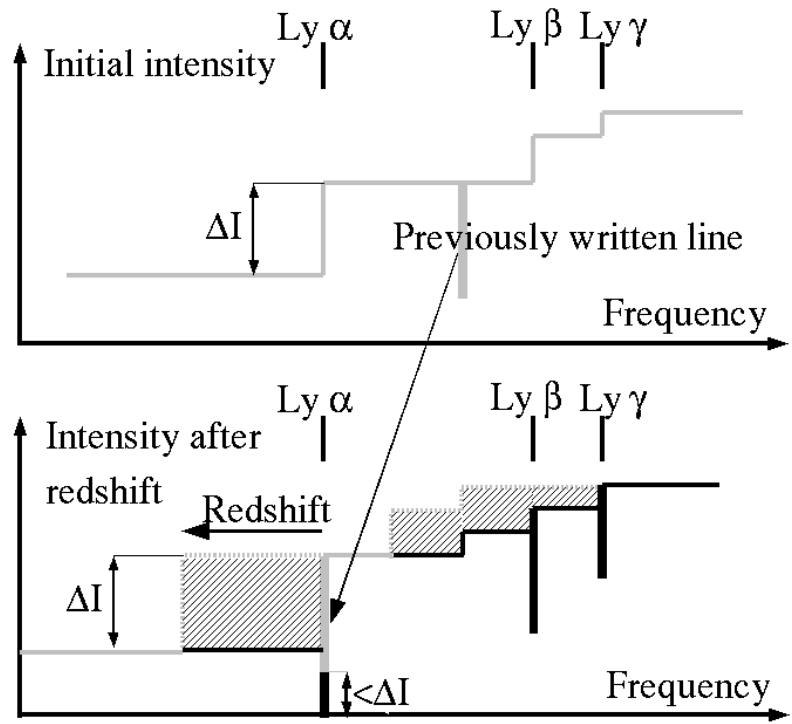} 
\end{center}
\caption{Multiplication of the Lyman spectral lines.}
\end{figure}
Fig. 8, top shows a continuous spectrum after an absorption of the main Lyman
lines, and an other absorbed line. During the redshift (fig. 8, low) , the
hachured regions are absorbed, but the intensity $\Delta I$ cannot be absorbed
when the previously written line comes on the Ly$_\alpha$ line. Therefore, the
redshifting stops until the absorption by the Ly$_\beta$ line is sufficient to
restart it. The absorption of the Ly$_\beta$ line must be larger than the
missing absorption of the Ly$_\alpha$ although the line is weaker because this
absorption does not produce a strong CREIL, the quadrupolar frequencies of the
$n=3$ level being lower than the frequencies of the $n=2$ level. Therefore,
during the stop, all lines, absorption or emission are strongly written into
the spectrum. Remark that this process works for a previously written emission
line because it produces an acceleration of the redshift, therefore a decrease
of the absorptions similar to an emission.

The process may be started by a line of an impurity, or by a Ly$_\beta$ line
written in a region where the redshift was forbidden. Then, the Lyman patterns
are linked, for instance by a coincidence of the Ly$_\beta$ line of a shifted
pattern with the Ly$_\alpha$ line of the gas. The characterisation of the
lines, for instance the Ly$_\beta$ line, by their frequencies may be replaced
by the redshift $z_{\beta,\alpha}$ needed to put their frequencies at the
frequency of the Ly$_\alpha$ line; thus, the linking of the lines gives:

$$z_{(\beta{\rm {resp.}\gamma,\alpha)}}=
\frac{\nu_{(\beta,{\rm resp.}\gamma)}-\nu_{\alpha}}{\nu_{\alpha}}
\approx\frac{1-1/(3^{2}{\rm resp.} 4^{2})-(1-1/2^{2})}{1-1/2^{2}}$$
$$z_{(\beta,\alpha)}\approx5/27\approx0.1852\approx3*0.0617;$$
$$z_{(\gamma,\alpha)}=1/4=0.025=4*0.0625.$$
Similar to 
$$z_{(\gamma,\beta)}\approx7/108\approx0.065.$$
 Notice that the resulting redshifts appear, within a good approximation, as
the products of $z_{b}=0.062$ and an integer $q$.
The intensities of the Lyman lines are decreasing functions of the final
principal quantum number $n$, so that the inscription of a pattern is better
for $q=3$ than for $q=4$ and \textit{a fortiori} for $q=1$.

\medskip
 Iterating, the coincidences of the shifted line frequencies with the Lyman
$\beta$ or $\alpha$ frequencies build a \textquotedblleft
tree\textquotedblright , final values of $q$ being sums of the basic values 4,
3 and 1. Each step being characterised by the value of q, a generation of
successive lines is characterised by successive values of $q : q_{1},q_{2}...$
As the final redshift is 
$$q_{F}*z_{b}=(q_{1}+q_{2}+...)*z_{b},$$
the addition $q_{F}=q_{1}+q_{2}+...$ is both a symbolic representation of the
successive elementary processes, and the result of these processes.
The metaphor \textquotedblleft tree\textquotedblright, is imprecise because
\textquotedblleft branches\textquotedblright  {}  of the tree may be
\textquotedblleft stacked\textquotedblright {} by coincidences of frequencies.
A remarkable coincidence happens for $q=10$, this number is obtained by the
effective coincidences deduced from an overlapping sequence of Lyman lines
corresponding to the symbolic additions:
$$10=3+3+4=3+4+3=4+3+3=3+3+3+1=... $$
$q=10$ is so remarkable that $z_{f}=10z_{b}=0.62$ may seem experimentally a
value of $z$ more fundamental than $z_{b}$.

\medskip
In these computations, the levels for a value of the principal quantum number
$n$ greater than 4 are neglected, for the simple reason that the corresponding
transitions are too weak.

 \section{Spectrum of the accreting neutron stars.}\label{quasar}
 \subsection{Building the spectrum}\label{constru}
The theory of the stars is very reliable, although some properties are not well
understood. The theory predicts that having lost a large part of its mass and
of its angular momentum, as it becomes a neutron star, an initially heavy star
may reach a step of its evolution in which it accretes the surrounding gas.
The fall of the gas heats the surface of the star so much that its temperature
is over 1 000 000 K. This temperature makes this {}``accretor'' so bright, in
particular at short wavelengths, that, in despite of its small size, it should
be easily observed \cite{Treves,Popov}. To solve this problem, study the
spectrum of these accretors (fig. 9).
\begin{figure} 
\includegraphics[height=10 cm] {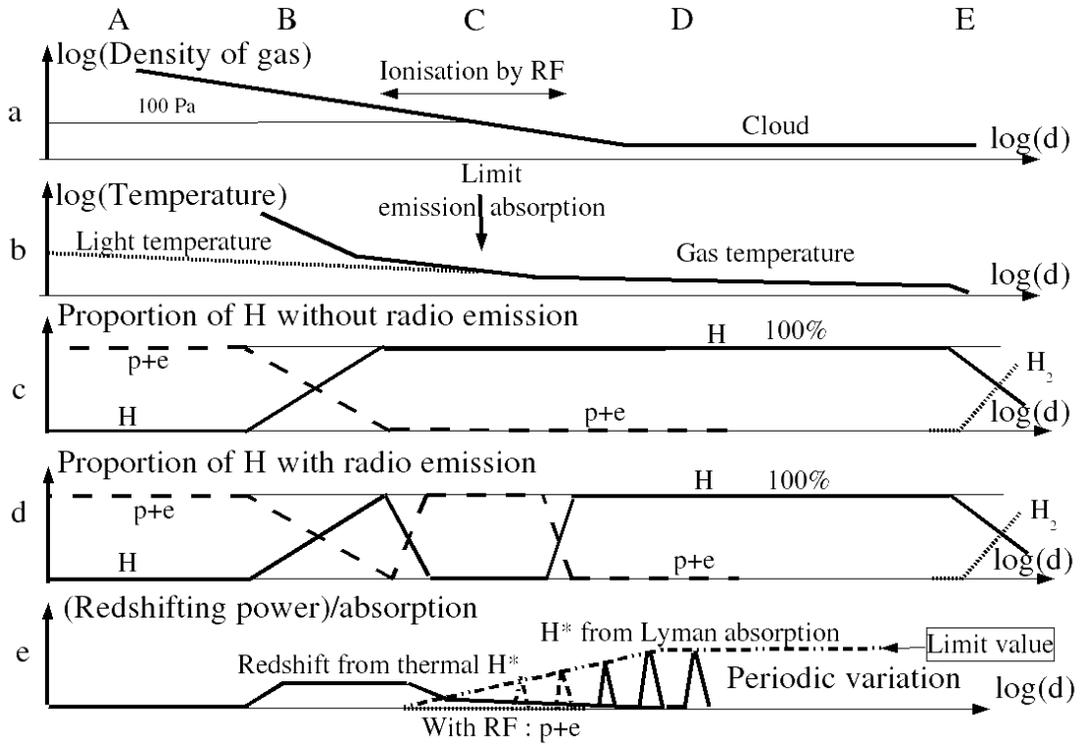} 
\caption{Building the spectrum of an accretor.}
\end{figure}

The graph {}``a'' of fig. 9 represents our rough hypothesis about the density
of gas around the star. The gas is far from an equilibrium because it falls
fast to the star. Therefore its density may change much slower than in the
hypothesis of an equilibrium, so that it may emit or absorb strongly light in
nearly constant conditions. We have written that the scales are logarithmic to
indicate that the scale of pressure is far from being linear : a millimetre
may represent less than a metre at the left, and a parsec at the right. At a
long distance, the density is supposed nearly constant, corresponding to a
cloud of gas. To make the theory, we suppose that the gas is nearly pure
hydrogen, the impurities being only able to produce emission or absorption
lines. We have indicated a region in which the density has the order of the
density in the discharge tubes, so that the gas may be ionised by radio
frequencies.

The graph {}``b'' represents our hypothesis about the temperatures of the gas
and the temperature of the light. The Planck's temperature of the light at a
point depends on the frequency, but we suppose that, except for spectral lines
it may be considered independent on the frequency at a point, decreasing with
the distance.

Very close to the star, the gas is strongly heated by its compression and by
electrons issued from lower regions. This heating stops quickly, so that it
remains only a radiative heating which stabilises relatively the temperature:
close to the star the temperature is high, a large quantity of hydrogen is
ionised into protons and electrons which do not absorb much energy, the
temperature drops fast. Very far, it does not remain much energy for a Lyman
absorption, the temperature drops, molecular hydrogen appears (graph {}``c'').

On graph {}``d'', we suppose that a strong radio emission ionises the gas so
that atomic hydrogen which appeared as the temperature decreased is destroyed
at pressures of the order of 100 Pa.

\medskip
Graph {}``e'' shows how the spectrum builds:

Very close to the star (column A), all atoms are ionised. The strong lines are
intense, therefore wide, but reabsorbed, so that the weak (forbidden) lines
may appear stronger and sharper. The fall of the gas adds a Doppler redshift
to the CREIL redshift which is locally slightly decreased by the Doppler
effect on the quadrupolar resonance.

In column B, atomic hydrogen appears, it is strongly excited by the collisions,
so that it redshifts the light. All lines are shifted while they are emitted,
they are so wide that they cannot be seen : there is a gap in the redshifts z.

In columns C-E, if there is no radio emission, the thermal excitation of atomic
hydrogen disappears, so that the periodicities described in the previous
section appear. At the beginning there are emissions, then absorptions. At
relatively high pressures, the hydrogen is quickly de-excited, so that the
absorptions are relatively strong, the lines are saturated. Close to the
centre of the lines, the saturation equilibrates the temperature of the light
with the temperature of the gas, so that the lines get the shape of a hat in
emission, of a trough in absorption. At these pressures, the gas is easily
ionised by radiofrequencies, so that these characteristic broad lines do not
appear.

At a longer distance, the lines become sharp. There is a large probability that
the atomic hydrogen disappears while the redshift is stopped, so that it is
not only the variations of redshifts, but the redshifts themselves that are
integer multiples of $z_b$.

\medskip
The previous description may be slightly changed, in particular because the
relation between the scales of density and temperature depends on the mass of
the star and the density of the cloud. 

 \subsection{Observation of the star}\label{etoile}
We have described a very complicated spectrum which is just the spectrum of a
quasar, explaining all its particularities:

* As the cloud was generated by an old star, it may contain heavy elements;

* Supposing that the relative frequency shifts $\Delta\nu/\nu$ are constant,
the
fine structure patterns are slightly distorted; the dispersion of the optical
constants in the CREIL shows that the hypothesis is not strict, so that it is
not necessary to suppose that the fine structure constant is a function of the
time \cite{Webb};

* There is a gap in the redshifts after the sharp emission lines \cite{Rauch};

* The broad lines which have the shape of troughs do no exist if there is a
strong radio emission \cite{Briggs,SAnderson};

* The observed periodicities \cite{Burbidge,Tifft,Hewitt,Bell,Comeau} are
simply produced by the propagation of the light in atomic hydrogen.

* A large part of the redshift is intrinsic, as found by Halton Arp \cite{Arp}.
Being not extraordinarily far, the quasars are not huge and powerful objects
\cite{Petitjean}.

\medskip
The building of so complicated a spectrum which requires so simple hypothesis,
and agrees so well with the observations is a proof that:

* The accretors are observed, but called quasars;

* The abundance of atomic hydrogen and the intensity of the CREIL are
sufficient
to produce strong intrinsic redshifts; is the {}``cosmological redshift''
produced by CREIL in the intergalactic space ?

 \section{Some other applications of the CREIL}\label{autres}

The origin of the observed redshifts may be split into CREIL, Doppler and
gravitational, the first one being generally the most important. Therefore,
the CREIL must be taken into account for most observations. The most
remarkable observations are:

A statistical over abundance of very red objects (VROs) is observed in close
proximity to quasars (Hall et al. \cite{Hall}, Wold et al. \cite{Wold}); in
particular, the galaxies which contain quasars are often severely reddened,
and redshifted relative to other galaxies having similar morphologies (Boller
\cite{Boller}). The quasar produces a CREIL redshift, providing far
ultraviolet radiation and maybe hydrogen around the VROs. 

* The bright and much redshifted objects seem surrounded by hot dust
\cite{Omont}, and it is difficult to explain the existence of this dust in
despite of the pressure of radiation and the abrasion by ions. The blueshift,
that is the heating of the thermal radiation by the CREIL is a simple
interpretation of the observations.

* Studying the variations of the frequency shifts on the Solar disk allows to
compute the fractions due to the Doppler effect and to the gravitation. It
remains a redshift proportional to the path of the light through the corona,
immediately explained by a CREIL effect.

* Radio signals were sent from the Earth to Pioneer 10 and 11, at a well
stabilised carrier frequency close to 2.11 GHz, and the Pioneers returned a
signal after a multiplication of the carrier frequency by 240/221. The
blueshift which remains after a standard elimination of the known frequency
shifts (Doppler, gravitation) is interpreted as produced by an
\textquotedblleft anomalous acceleration\textquotedblright {}(Anderson et al.
\cite{Anderson}). The CREIL allows to preserve celestial mechanics : Assume
that the solar plasma between these Pioneer probes and the Earth contains
molecules possessing resonances in the megaherz range (either or for instance
Lyman pumped atomic hydrogen). These molecules transfer energy from the solar
radiation not only to the thermal radiation but to the radio signals too :
Planck's temperature of the radio signals is higher than 2.7K to allow a
detection, but much lower than the temperature of the solar radiation. The
CREIL requires an incoherence, that is a high frequency modulation of the
light. The emission of the Pioneers is very weak, quickly mixed with the
thermal noise which provides a modulation. Crucial experiments could be done,
studying the variation of the frequency shift as a function of the modulation,
either changed by a variation of the intensity of the carrier, or changed by a
variable, known modulation.

* V. A. Kotov and V. M. Lyuty \cite{Lyuty, Kotov} observed oscillations of the
luminosity of stars and quasars with a period of 160,01 mn. While the light is
redshifted, this period is not. Using CREIL, it is clear that the light pulses
are redshifted, but that their starts are not subject to a frequency shift
\cite{Lempel}. On the contrary, supposing a change in the scale of time by an
expansion of the universe, this result cannot be explained. Therefore,
thinking that the observations are reliable, there is no expansion of the
universe.

\section{Conclusion}
Avoiding the use of the CREIL, an elementary optical parametric effect, is
never justified by the supporters of the big bang.

Using this effect to study the spectrum of an accreting neutron star shows a
very complicated spectrum which appears being just a spectrum of quasar. It
cannot be a coincidence, so that the \textquotedblleft intrinsic
redshifts\textquotedblright {} are surely produced by the CREIL.

The lack of redshift of the variations of luminosity observed in stars and
quasars with a period of 160 minutes shows that the \textquotedblleft
cosmological redshift\textquotedblright {} is produced by a CREIL effect, so
that there is no expansion of the Universe

\end{document}